\documentclass[preprint,12pt]{elsarticle}

\usepackage{amssymb}
\usepackage{amsmath}
\usepackage{graphicx}
\usepackage{subcaption}
\usepackage{xcolor}

\journal{Chemical Engineering Science}

\begin{document}

\begin{frontmatter}



\title{Tortuosity in the Brick and Mortar model based on Chemical conduction}


\author{Bj\"orn Stenqvist}
\ead{bjorn.stenqvist@teokem.lu.se}
\author{Emma Sparr}
\address{Division of Physical Chemistry, Lund University, POB 124, SE-22100 Lund, Sweden}

\begin{abstract}
Diffusion is a reoccurring phenomena in many fields and is affected by the geometry in which it takes place. Here we investigate the effects of geometry on diffusion in a Brick and Mortar model system. The tortuous effects are evaluated based on generalized Fick's law, i.e. diffusion driven by differences in chemical potential. The presented formalism gives a general (semi-)exact analytic expression for the tortuosity using impermeable Bricks, which is successfully validated against standard techniques and finite element method results. The approach allows for anisotropic properties of the Mortar, which we show can be significant and is not captured with known analytic techniques. Based on the introduced concept of chemical conductivity we also find generalized Fick's law consistent with Ohm's and Fourier's law in terms of their constituent parts, which further makes the main results for brick and mortar structures directly applicable to diffusion of either charge, heat, or mass.
\end{abstract}



\begin{keyword}
Tortuosity; Fick's law; Fourier's law; Ohm's law; Brick and Mortar


\end{keyword}

\end{frontmatter}


\section{Introduction}
Diffusion is a phenomena that occurs in many different forms, both in nature and technology. For example, diffusion of charges in electrical circuits, energy in meteorology, and mass diffusion in molecular chemistry. Here our main focus is on the latter which typically is described by Fick's law. The driving-force for mass diffusion is usually described in term of the concentration-gradient. This approach is however only valid for homogeneous mediums and ideal scenarios, i.e. unit activity coefficient. A broader representation including discontinuous concentration profiles and non-ideal cases is a \emph{generalized} Fick's law in which diffusion is driven by differences in chemical potential of the diffusing compound\cite{Onsager1932Irreversible}. This approach is in essence more accurate whilst convenient to use in inhomogeneous systems. As the chemical potential is smooth in any system its gradient is continuous, contrary to the gradient of the concentration profile which might be discontinuous. Such an example is the interfaces between air and a condensed phase, where it is not always meaningful to discuss a concentration gradient. Another example is found in heterogeneous media where the solubility of the diffusing substance can very substantially between different regions. Such systems are found in many complex materials in biology and in technical applications. 

In this work we interpret the combination of constituent parts of the generalized Fick's law at steady-state and compare to Ohm's law of charge conduction and Fourier's law of thermal conduction. The analogy between these laws is by no means new, however the presented interpretation of the transport coefficients and specifically the here introduced concept of \emph{chemical conductivity} is. This entity makes the three laws consistent in term of physical parameters, and their common form provides a general phenomenological understanding. By then using this framework we find an analytic formula for the tortuosity in the Brick and Mortar model which is applicable for electric, thermal or mass diffusion. The result can be used to describe flux in barrier materials\cite{dou2015transparent}, nacre micro-structures\cite{yao2010artificial} and nanocomposite membranes \cite{Chai2014}, or cellulose nanofiber foams design to sustain drug delivery systems\cite{Svagan2016solid}. As the shape of the named laws are also similar to that of Darcy's law for hydrodynamic flow\cite{Whitaker1986} and Newton's law of
viscosity\cite{OCHOATAPIA20071} the general results in this paper is also valid for such systems.

The derived expression for the tortuosity is benchmarked on typical stratum corneum (SC) geometries, the upper layer of the skin, for which the Brick and Mortar model is commonly used\cite{Heisig1996,Hadgraft2011,Walicka2018}. The Bricks then represent corneocytes and the Mortar represent lipid layers. Corneocytes are dead cells filled with solid keratin rods, which can swell significantly in water. The corneocytes are embedded in a multilayer lipid matrix where the lipid layers are on average arranged parallel to the skin surface. These extracellular lipids makes up the only continuous route through SC\cite{Heisig1996}. In case of hydrophobic diffusing compounds, the corneocytes can be considered impermeable and SC can thereby be seen as a Brick and Mortar system with impermeable bricks. Furthermore, the extracellular lipids constitute an anisotropic matrix where one expect diffusion to differ in the directions parallel and perpendicular to the bilayer normal in the multilayer structure. The here presented approach is able to capture inhomogeneities in the lipid matrix which for SC is an important aspect to consideration\cite{bouwstra1991structural}. This novel property of the method makes it forthright to study anisotropic features of SC lipids using the Brick and Mortar model, contrary to standard analytic methods\cite{cussler1988barrier,johnson1997evaluation}.

\section{Theory}
Generally the flux $\boldsymbol{j}$ in a diffusion process can be described\cite{Griffiths1999Introduction} as a product between the conductivity $\sigma$ and the acting force $\boldsymbol{f}$,
\begin{equation}\label{eq:flux_gen}
  \boldsymbol{j} = \sigma \boldsymbol{f}.
\end{equation}
Any such conservative force can be described as the (negative) gradient of a potential, e.g. electric, gravitational, or chemical. By utilizing this feature it is possible to derive among others Ohm's, Fourier's and Fick's law.\cite{Onsager1931Reciprocal,Onsager1953Fluctuations} A generalized Fick's law can therefore be described by a gradient of the chemical potential \cite{Onsager1932Irreversible} after which its concentration depending counterpart is retrieved in the special case of ideal conditions\cite{crank1956mathematics}. The mass flux $\boldsymbol{j}$ [kg/m$^2$s] can be described by\cite{Hearon1950}
\begin{equation}\label{eq:gen_ficks}
  \boldsymbol{j} = -Uc\boldsymbol{\nabla}\mu
\end{equation}
where $U$ [mol s/kg] is the mass mobility defined as the velocity per unit force [m/s / J/mol m], $c$ [kg/m$^3$] the concentration, and $\mu$ [J/mol] the chemical potential. Another version of the generalized Fick's law\cite{evans1999colloidal} based on the (ideally defined) diffusion coefficient $D_0$ [m$^2$/s] can furthermore be derived from Eq.~\ref{eq:gen_ficks}. Note that for ideal systems $D_0=URT$, where $R$ [J/K mol] is the gas constant and $T$ [K] the temperature, which is commonly denoted as the Einstein relation. For an overview of the relationship between mobility, diffusion coefficient, and gradient of chemical potential, and Fick's and Ohm's laws we refer elsewhere\cite{Kocherginsky2016}. By comparing the above equation to Eq.~\ref{eq:flux_gen} we recognise $U$ and $c$ to be constitute parts of the conductivity, which we further define as the \emph{chemical} conductivity $\sigma_c$ and get 
\begin{equation}\label{eq:gen_ficks_cc}
  \boldsymbol{j} = -\sigma_c\boldsymbol{\nabla}\mu.
\end{equation}
Here, the introduction and definition of the concept chemical conduction and its analogy with other diffusion laws is to our knowledge noval, and further details on this matter can be found in the supplementary information (SI) Sec.~S1. Conductivity as a general concept is well-known and thus provides an intuitive understanding whilst it condenses and simplifies the transport equation. In Sec.~S2 in the SI we also derive an alternative form of the generalized Fick's law based on an activity coefficient, which might sometimes be more practical to use. 

Based on the generalized Fick's law and the chemical conduction we in the following section analytically solve diffusion in one dimension. This solution is then used to define the effective resistance of a Brick and Mortar membrane, which is sufficient to retrieve the tortuosity of said membrane. Finally we analyze results based on the derived expression, and also compare to existing analytic methods and finite element method results. 

\subsection{Fick's law in 1D}
The generalized Fick's law, see Eq.~\ref{eq:gen_ficks_cc} or Eq.~S4 (SI) for the ideally defined variant, in one dimension (1D) using known boundary conditions, and a chemical conductivity as a function of the single coordinate $z$, is enough to retrieve a solution in steady-state. We get
\begin{equation}\label{eq:fick_sol}
\mu(z) = \mu(z_0) + \lambda(z)\Delta\mu
\end{equation}
where $z_0$ and $z_f$ defines the boundaries for the $z$-coordinate, $\Delta\mu=\mu(z_f) - \mu(z_0)$,
\begin{equation}
\lambda(z) = \frac{\int_{z_0}^z\sigma_c^{-1}(z^{\prime})dz^{\prime}}{\int_{z_0}^{z_f}\sigma_c^{-1}(z^{\prime})dz^{\prime}},
\end{equation}
and finally the flux as
\begin{equation}\label{eq:gen_flux}
j = -\frac{\Delta\mu}{\int_{z_0}^{z_f}\sigma_c^{-1}(z)dz}
\end{equation}
which has also been found in an earlier work\cite{kocherginsky2010mass}.

\subsection{Geometry and properties}
\begin{figure}[ht]
\centering
   \includegraphics[width=0.8\columnwidth]{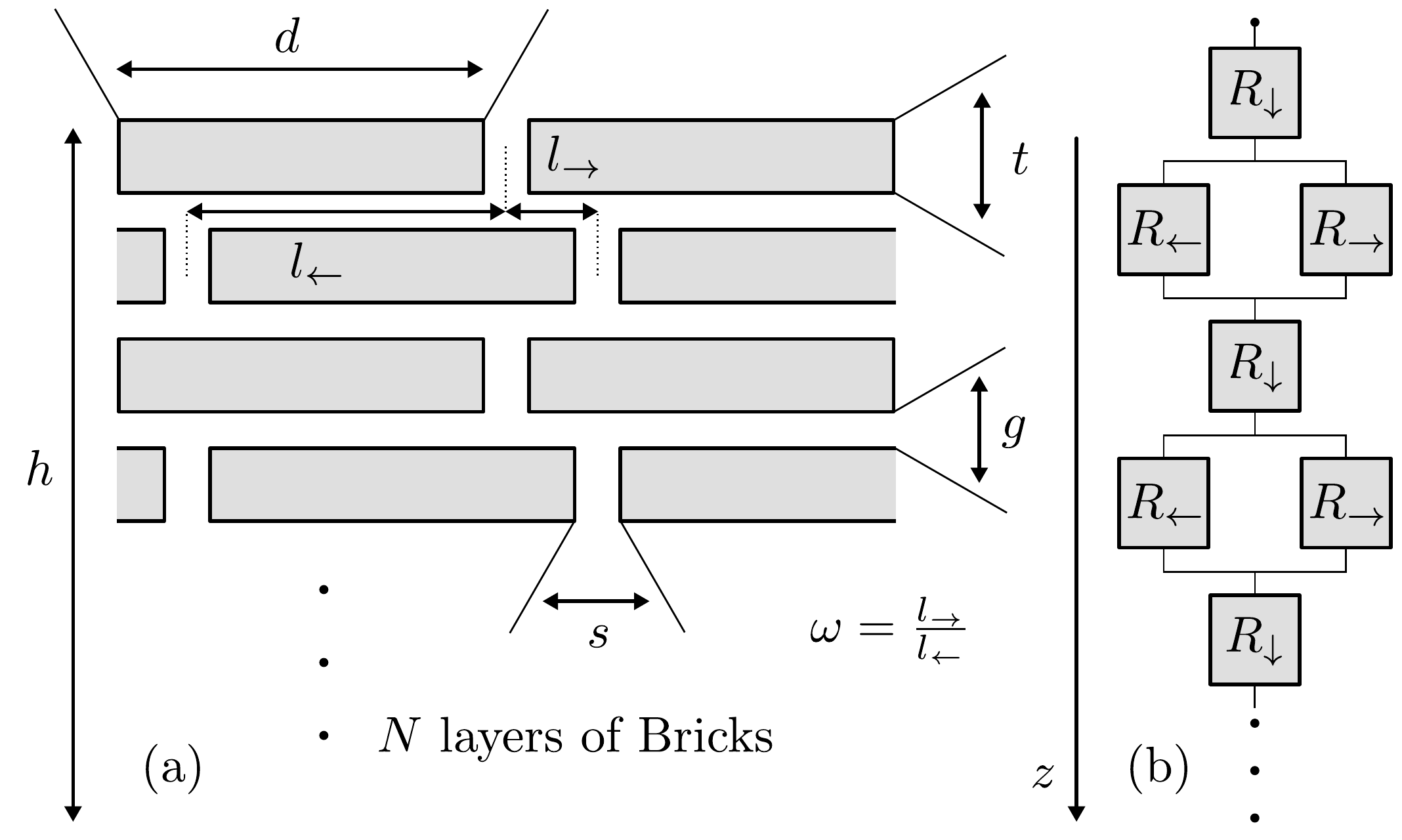}
\caption{(a) Geometry of the Brick and Mortar model. Dashed lines are centered in their respective slit. (b) Circuit representing the total resistance for the Brick and Mortar model where $R_{\downarrow}$ is the resistance for transverse diffusion in the Mortar, and $R_{\leftarrow}$ and $R_{\rightarrow}$ represent resistance for lateral diffusion in the Mortar. Here the Bricks are considered as impermeable. }
\label{fig:bm_model} 
\end{figure}

In this section we will introduce the geometry of the Brick and Mortar membrane, see Fig.~\ref{fig:bm_model}a, and its properties. The geometric variables: lateral period $L$, Brick width $d$, lateral spacing between Bricks $s$, vertical spacing between Bricks $g$, Brick thickness $t$, total membrane thickness $h$, ``left'' offset slit-distance $l_{\leftarrow}$, and ``right'' offset slit-distance  $l_{\rightarrow}$, all in units of length [m] are all illustrated in this figure. Furthermore, the offset ratio $\omega=l_{\rightarrow}/l_{\leftarrow}$ which is a measure of the overlap of the Bricks, and the number of layers of Bricks $N$, both unitless. For consistency we have kept similar notation as in previous models of Brick and Mortar systems\cite{johnson1997evaluation}. Throughout this work we assume the Bricks to be \emph{impermeable}. Therefore any particle diffusing through the membrane has to travel in the Mortar matrix. The Bricks and Mortar system is laterally periodic and thus the lateral crossing-points are indistinguishable from one-another. By crossing-point we mean a point where lateral and transverse mortar diffusing-paths meet. Without loss of generality we from now on assume the period $L=d+s$ to be constant. If not stated otherwise we assume $\sigma_{\parallel}$ as the (constant) chemical conductivity for lateral diffusion and $\sigma_{\perp}$ as the (constant) chemical conductivity for transverse diffusion. To assume constant chemical conduction (for ideal systems) has been used in previous work.\cite{Wakeham1979} The presented procedure is however readily generalizable also for non-constant chemical conductivities, see Sec.~S3 in the SI. 

\subsection{Resistance of a membrane}
Now we derive an expression for the effective chemical resistance of a Brick and Mortar membrane. To put it simply, starting from the upper boundary, a particle will start to move transversely until it reaches a crossing-point where it will then move laterally, either to the left or right. Either way, the next crossing-points are indistinguishable due to periodicityty. Then, again, the particle will travel transversely and this procedure repeats until the lower boundary is reached. This pattern can be easily illustrated by a circuit diagram, see Fig.~\ref{fig:bm_model}b. The derivation of the effective chemical resistance will therefore mirror the derivation for an effective electrical resistance assuming 1D diffusion.

We start by denoting the mass flow rate as $\dot{m}$ [kg/s] and note that $\dot{m} = -j_cA$ where $A$ [m$^2$] is the cross-area through which the mass flow. Analogously to electric theory we further define the chemical resistance $R$ [Js/kg] as
\begin{equation}\label{eq:resistance}
R \equiv \frac{\Delta\mu}{\dot{m}}.
\end{equation}
The total resistance for the Brick and Mortar membrane is then the sum of all resistances (effective or otherwise) which are connected in series. 
Since Fick's law is solved in 1D separately for each subsection of the path (i.e. over each resistor), we need to assume uniform properties over the width of the diffusion channel, in our case $g$ and $s$. See Sec.~S4 in the SI for more discussion on this assumption. Thus the cross-area of diffusion will be approximated by $L_Dg$/$L_Ds$, where $L_D$ [m] is the depth (going inward in Fig.~\ref{fig:bm_model}a). Note that as $g\to 0$ and $s\to 0$, i.e. small lateral spacing between the Bricks, the approximation is exact. Still it might be exact for larger $g$/$s$. By using the the mass flow rate, Eq.~\ref{eq:resistance}, the respective solutions to the flux in Eq.~\ref{eq:gen_flux}, and the expressions for serial and parallel effective resistance we then get
the total resistance for a membrane consisting of $N$ layers of Bricks as
\begin{equation}
R_N = \frac{h}{Ls\sigma_{\perp}} + (N-1)\frac{l_ll_r}{LgP\sigma_{\parallel}}.
\end{equation}
As a state of reference we for the following presentation need to retrieve the resistance of a homogeneous membrane without Bricks, i.e. $s=L$ and $l_{\leftarrow}=0$ (or equivalently $l_{\rightarrow}=0$), which is $R_0 = h/L_DL\sigma_{\perp}$.

\subsection{Tortuosity}
By using the results in the previous section, we find that the tortuosity, defined as $\tau = j_0/j_N$ transforms to $\tau = R_N/R_0$. After introducing the offset ratio $\omega=l_{\rightarrow}/l_{\leftarrow}$ this can be simplified to
\begin{equation}\label{eq:tau_final}
\tau_{\mu} = 1+\frac{d}{s} + \frac{L^2\omega}{g\left(g+t\frac{N}{N-1}\right)(1+\omega)^2}\frac{\sigma_{\perp}}{\sigma_{\parallel}}
\end{equation}
where subscript $\mu$ indicate derivation based on the gradient of the chemical potential. Note that Eq.~\ref{eq:tau_final} is exact and without assumptions for $\omega=\{0,\infty\}$, i.e. $l_{\leftarrow}=0$ or equivalently $l_{\rightarrow}=0$, as for these cases there are no lateral flux. Finally, using a single layer of Bricks $N=1$, Eq.~\ref{eq:tau_final} breaks down and instead we have $\tau_{\mu} = 1 + d/s$.

\subsection{Existing models in the literature}
There exists many approaches describing the geometrical impact on Brick and Mortar membrane permeability \cite{Michaels1975Drug,Cussler1990,LANGELIECKFELDT1992183,johnson1997evaluation,KUSHNER20073236} and here we compare to one of the more rigorous and flexible models.\cite{cussler1988barrier,johnson1997evaluation} Several of the mentioned approaches are valid within their own regime but in this work we aim for a general and diverse formula. In the next section we will therefore compare our results to the previously  published formula
\begin{equation}\label{eq:tauj_final}
\tau_c = 1 + \frac{2g}{h}\ln\left(\frac{L}{2s} \right) + \frac{NLt}{sh} + \left(N-1\right)\frac{L^2}{hg}\frac{\omega}{(1+\omega)^2}
\end{equation}
which like the introduced approach is based on one-dimensional diffusion\cite{cussler1988barrier,johnson1997evaluation}. The subscript $c$ in Eq.~\ref{eq:tauj_final} indicate derivation based on gradient of the concentration. In the original work\cite{johnson1997evaluation} the width of the Bricks $d$ is assumed to be equal to the period $L$, yet the approach is straight-forward to adjust to geometries when $d\ne L$. The above equation has thus been generalized accordingly. The derivation of Eq.~\ref{eq:tauj_final} is valid whenever $s\ll L \ll 2h$. That is, the lateral spacing between the Bricks needs to be much smaller than the period, and the period needs to be much smaller than the total thickness of the membrane. For many systems these assumptions are not valid, where one example is SC which for typical geometrical values gives $L\approx 2h$. We also note an interesting, although erroneous, feature of Eq.~\ref{eq:tauj_final}: As $\omega\to 0$ the tortuosity still depends on $g$. However, from symmetry and steady-state we acknowledge no lateral flux and thus the choice of $g$ should be irrelevant. Similar flawed conclusions can however be drawn based on many existing models and is thus not unique. Finally we want to highlight that no previous approach for the tortuosity captures anisotropic features of the Mortar, which indeed our model does.

\section{Results and Discussion}
In this section we assess the tortuosity based on the derived $\tau_{\mu}$, see Eq.~\ref{eq:tau_final}, for different Brick and Mortar geometries and properties. After an initial analysis we then compare to results using existing models $\tau_c$, see Eq.~\ref{eq:tauj_final}. The results from $\tau_{\mu}$ are depicted by solid lines and $\tau_c$ by dashed lines.

\begin{figure}[!t]
    \centering
    \begin{subfigure}[b]{0.4\columnwidth}
        \includegraphics[width=\textwidth]{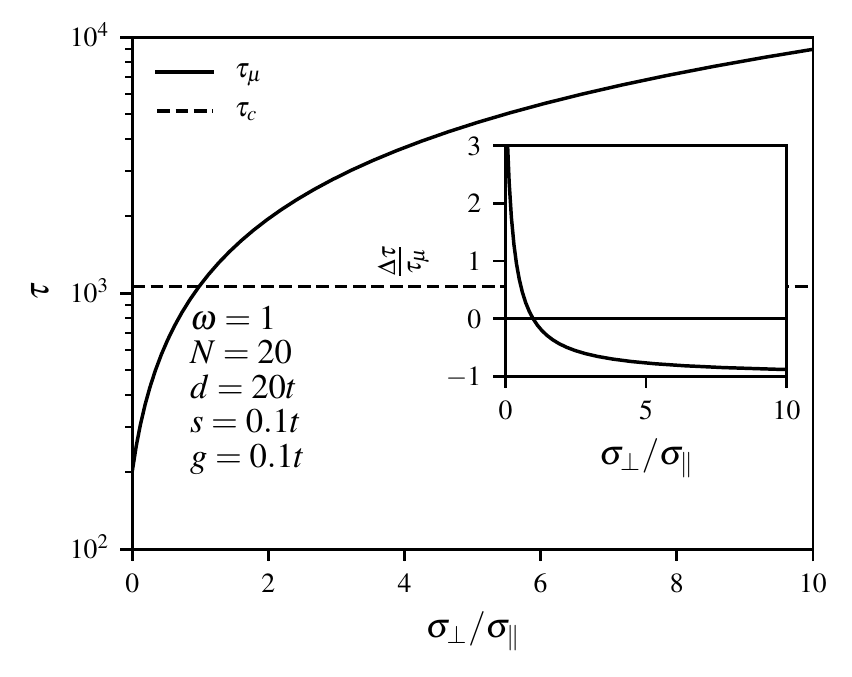}
        \caption{}
        \label{fig:tau_newdata_q}
    \end{subfigure}
    ~ 
    
    \begin{subfigure}[b]{0.4\columnwidth}
        \includegraphics[width=\textwidth]{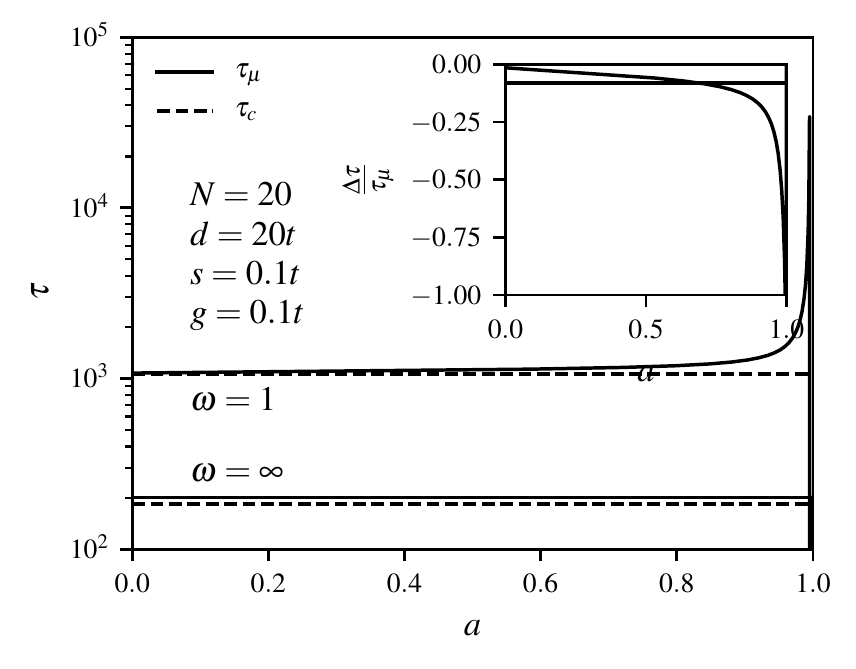}
        \caption{}
        \label{fig:tau_newdata_a}
    \end{subfigure}
    ~ 
    \begin{subfigure}[b]{0.4\columnwidth}
        \includegraphics[width=\textwidth]{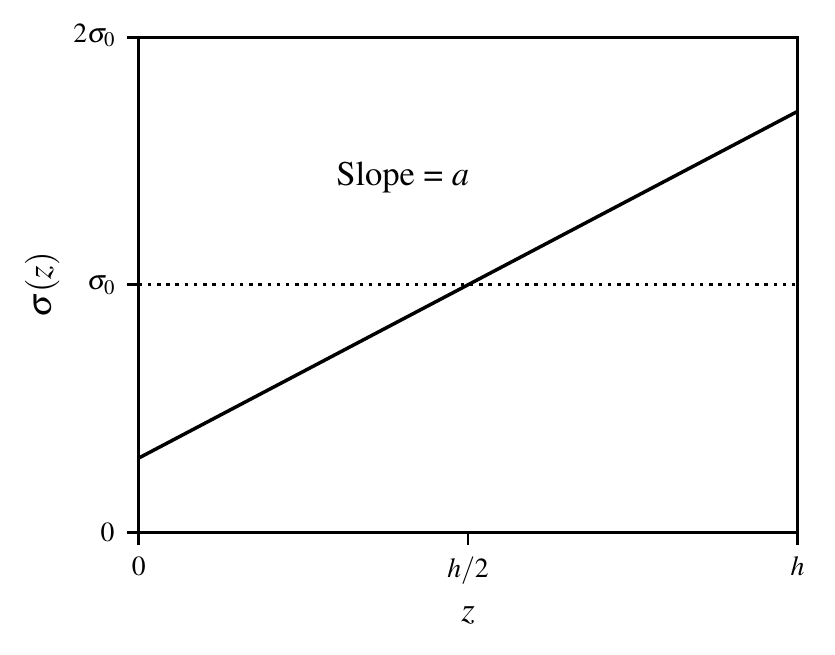}
        \caption{}
        \label{fig:tau_compare_a_comp}
    \end{subfigure}
    \caption{Tortuosity as a functions of (a) $\sigma_{\perp}/\sigma_{\parallel}$, and (b) $a$, see Eq.~S23 in SI for details. The insets show the difference between the two approaches divided by this work. (c) Illustrate the chemical conductivity dependence of $a$. }\label{fig:tau_newdata}
\end{figure}

We start by investigate cases which current standard techniques (e.g. Eq.~\ref{eq:tauj_final}) for the tortuosity\cite{cussler1988barrier,johnson1997evaluation} can not cover, yet which is straight-forward using the approach in this work. In Fig.~\ref{fig:tau_newdata_q} we investigate the impact of different lateral and transverse chemical conductivities, i.e. highly anisotropic properties relevant to many physical systems. We include results based on $\tau_c$, which assumes an isotropic mortar, to contrast the results of $\tau_{\mu}$. As $\sigma_{\perp}/\sigma_{\parallel}\to 0$, transverse transport is the main barrier for diffusion and thus in this limit $\tau_{\mu}$ is independent of $\omega$. The tortuosity increases linearly with $\sigma_{\perp}/\sigma_{\parallel}$, see Eq.~\ref{eq:tau_final}, and thus the $\omega$ dependence becomes increasingly important. At the intersect between the two curves, i.e. at $\sigma_{\perp}/\sigma_{\parallel}=1$ which represent a truly homogeneous mortar, the slope of $\tau_{\mu}$ is fairly steep. Consequently, a small change in chemical conductivity ratio at this point will induce a large change in the tortuosity. The impact is striking and should not to be ignored for non-extreme values $0 \ll \omega \ll \infty$. Furthermore Fig.~\ref{fig:tau_newdata_a} show the results when the chemical conductivity is instead defined as a linear function of the $z$-position. Here the slope is denoted $a$, see Sec.~S5 in the SI for more details on the used expressions. For a fairly homogeneous chemical conductivity (i.e. small $a$) the slope of the $\omega=1$ curve is roughly flat and we see no large effect. At larger $a$, in this case $a \sim 0.8$, however the tortuosity increases steeply. This indicate that a medium with low to moderate gradient in chemical conductivity can be modelled accurately using a homogeneous approximation. On the contrary, for an alternating gradient in chemical conductivity, as we saw in Fig.~\ref{fig:tau_newdata_q}, such an approximation is not valid.

\begin{figure}[!t]
    \centering
    \begin{subfigure}[t]{0.4\columnwidth}
    \centering
        \includegraphics[width=\textwidth]{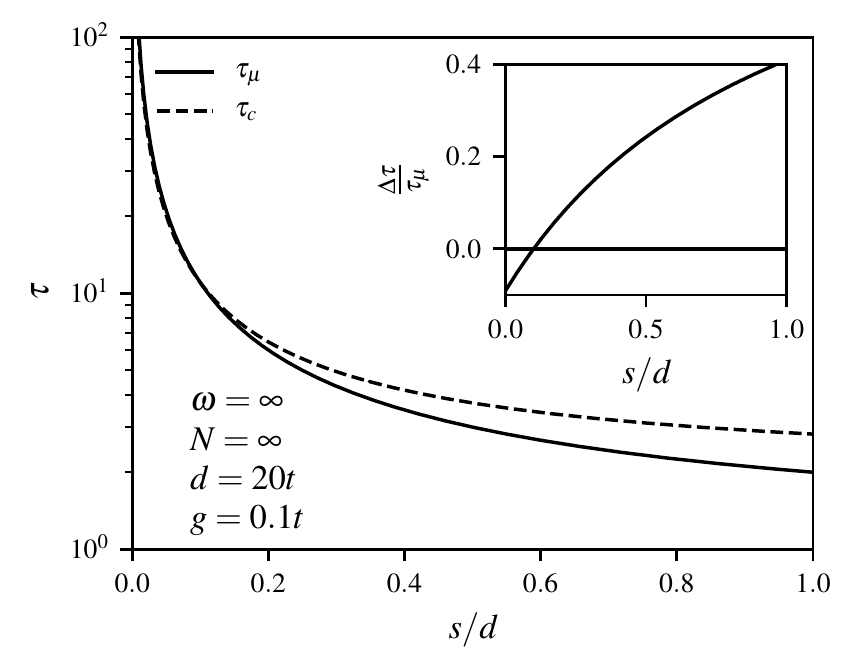}
        \caption{}
        \label{fig:tau_compare_s}
    \end{subfigure}
    ~ 
    \begin{subfigure}[t]{0.4\columnwidth}
    \centering
        \includegraphics[width=\textwidth]{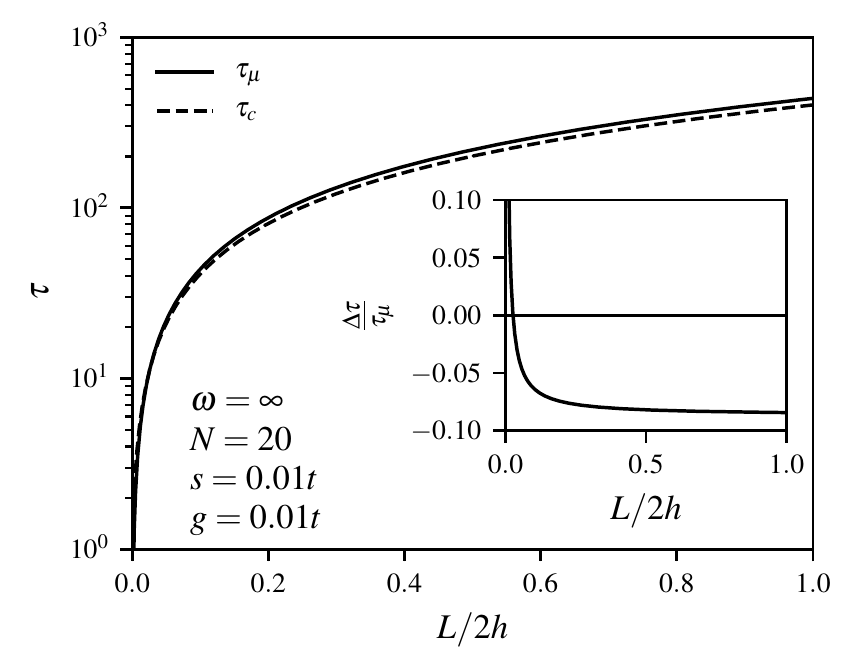}
        \caption{}
        \label{fig:tau_compare_d}
    \end{subfigure}
    ~ 
    \begin{subfigure}[t]{0.4\columnwidth}
    \centering
        \includegraphics[width=\textwidth]{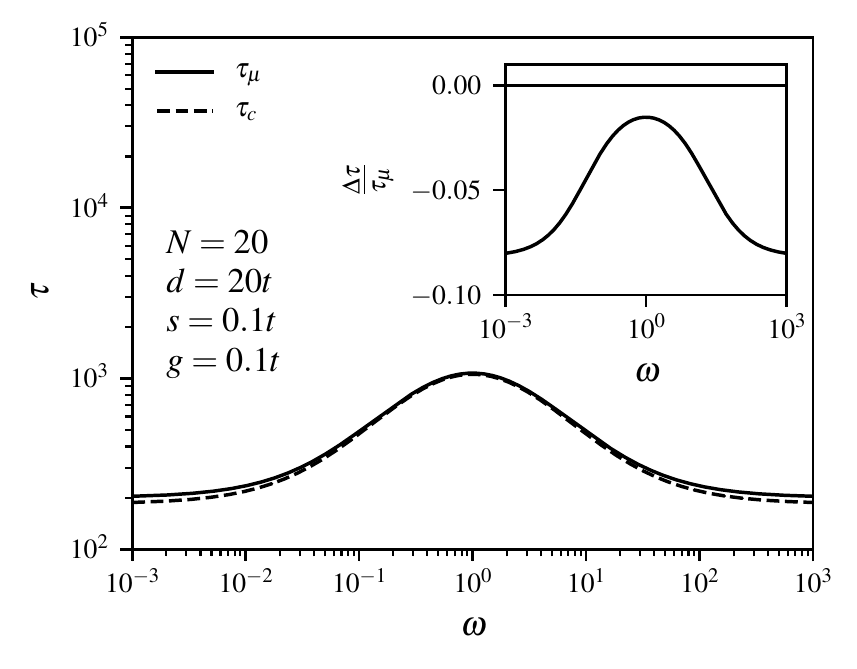}
        \caption{}
        \label{fig:tau_compare_o}
    \end{subfigure}
    \caption{Tortuosity as a functions of (a) $s/d$ where $\tau_c$ is valid for $s/d\ll 1$ and $\tau_{\mu}$ for small $s$, (b) $L/2h$ where $\tau_c$ is valid for $L/2h\ll 1$, and (c) $\omega$ where $\tau_{\mu}$ is exact for $\omega=\infty$ (or equivalently $\omega=0$). The insets show the difference between the two approaches $\Delta\tau=\tau_c-\tau_{\mu}$ divided by $\tau_{\mu}$. Note in (b) the lower limit of $L/2h=438^{-1}$ for $\tau_{\mu}$ at which $d=0$ and we thus have a homogeneous membrane -- that is unit tortuosity.}\label{fig:tau_compare}
\end{figure}

We now move on to analyze results in systems with isotropic mortar, i.e. systems where $\tau_c$ can also be applied. In Fig.~\ref{fig:tau_compare_s} we present results for an infinitely layered membrane ($N=\infty$) where bricks are perfectly aligned ($\omega=\infty$). The tortuosity tends to infinity as $s/d\to 0$ which is reasonable since at $s=0$ there is no transverse diffusion-path, making diffusion through the membrane impossible. At large $s/d$ the membrane asymptotically equals a homogeneous membrane of only mortar, giving a unit tortuosity -- that is maximum flux. Fig.~\ref{fig:tau_compare_d} illustrates how the tortuosity increases as the thickness of the full membrane decrease (note constant $L$). Effectively this means that as the relative thickness of the Bricks $t$ decreases compared to $L$, the tortuosity increases and finally at $L/2h=\infty$ we get $\tau_{\mu}=\infty$. Fig.~\ref{fig:tau_compare_o} shows the tortuosity as a function of the offset ratio $\omega$. At $\omega=1$ the tortuosity is at a maxima, which is straight-forward to acknowledge from the derivatives of $\tau_{\mu}$ with respect to $\omega$ or simply due to symmetry. For $\omega\to 0$ (or $\omega\to \infty$) the tortuosity only depends on the ration $d/s$ as the Bricks are perfectly aligned which in steady-state makes for a truly one-dimensional problem since from symmetry there is no lateral flux.

Whilst comparing $\tau_{\mu}$ and $\tau_c$ we acknowledge that the here presented tortuosity is (semi-)exact in Fig.~\ref{fig:tau_compare_s} and Fig.~\ref{fig:tau_compare_d} (i.e. exact at least for small $g$ and $s$), and indeed exact in the extremities of Fig.~\ref{fig:tau_compare_o}. Therefore, we argue most differences to be due to the approximations inherent to $\tau_c$. In Fig.~\ref{fig:tau_compare_s} the inset show differences between $-10\%$ to $+40\%$ pending $s/d$, which is substantial. The differences in Fig.~\ref{fig:tau_compare_d} are also considerable yet seems to converge towards $\sim 8\%$ as $L/2h$ increases, however, as $L/2h\to 438^{-1}$ (the lower limit) the difference is $\sim 90\%$. So far we have compared $\tau_{\mu}$ and $\tau_c$ when the former is (semi-)exact, yet now we turn to Fig.~\ref{fig:tau_compare_o} where $\tau_{\mu}$ is truly exact at $\omega=0$ and $\omega=\infty$. Here the difference for any value of $\omega$ is less than $9\%$, with a minima around $\omega=1$ at $\sim 2\%$. Since the maximal deviation between the two models are at the extremities where $\tau_{\mu}$ is exact, and thus at intermediate $\omega$ the differences are comparable small, we conjecture $\tau_{\mu}$ to be more accurate than $\tau_c$ on the interval as a whole. For more details, see Sec.~S6 in the SI where we expand on the comparison between $\tau_{\mu}$ and $\tau_c$. Finally we in Sec.~S7 present a similar case to Fig.~\ref{fig:tau_compare_s} but with vertically stretched Bricks, with similar tortuosity.

\begin{figure}[!ht]
  \caption{Tortuosity compared to finite-element model results. Data was extracted from Fig.~6b and $\tau_{c^*}$ is described in Eq.~6 in reference\cite{frasch2003steady}. Comparison between $\tau_{\mu}$ and $\tau_{\rm FEM}$ is made indirectly through their relationship to $\tau_{c^*}$. The used parameters was $\omega\in\{1,3,8,19,38,\infty\}$, $d\in\{20t,40t,60t\}$, $g=s=0.1t$, and $N=20$. The black line symbolizes a one-to-one relationship between the axes. }
  \centering
    \includegraphics[width=0.8\columnwidth]{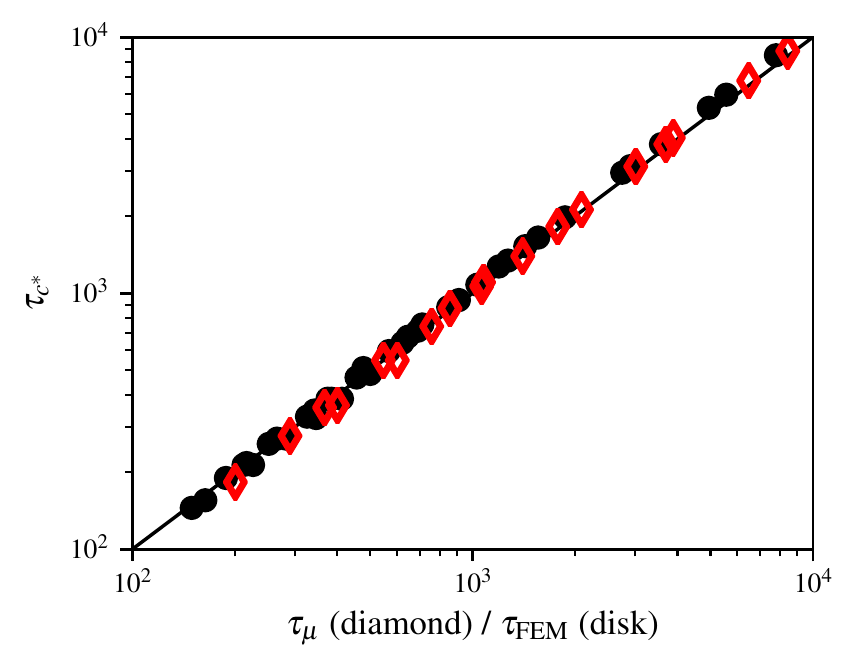}
    \label{fig:FEM}
\end{figure}

Lastly, in Fig.~\ref{fig:FEM} we present a comparison between our method and finite element method (FEM) results. Overlap between disks and diamonds in Fig.~\ref{fig:FEM} represent agreement between FEM and our results, which we generally have.

\subsection{Implications for Stratum corneum}
In its current 1D form, the presented model is foremost applicable to describe diffusion of hydrophobic substances in SC. This as the corneocytes (Bricks) then constitutes the dominant obstacles with low permeability, and diffusion primarily transpires in the surrounding lipid matrix. Unlike preceding methods\cite{cussler1988barrier,johnson1997evaluation} the here given model can capture anisotopic characteristics of this lamellar extracellular lipid phase, which is a significant trait of SC extracellular lipids\cite{bouwstra1991structural}. Especially transport properties such as mass mobilities and diffusion coefficients are expected to be of large importance to diffusion as (generally) particles do not partition equally between the polar headgroups and apolar hydrocarbon chains\cite{evans1999colloidal}. The introduced method is exact as $g\to 0$ and $s\to 0$ which generally are accurate assumptions for SC\cite{edwards1994linear,johnson1997evaluation}. Dissimilar to prior works\cite{cussler1988barrier,johnson1997evaluation} the presented model does not suffer from the for SC invalid assumption that $L\ll 2h$. That is, there is no assumption that the width of the corneocytes (Bricks) are much smaller than the total thickness of SC. Furthermore, other approaches\cite{KUSHNER20073236} suffers from inherent fallibility's as for example $g=s=0$ gives finite tortuosity.

The presented approach can also be based on hexagonal prism Bricks instead of the cuboids, which makes the membrane even more similar to SC structures\cite{edwards1994linear,yu2018geometrical}. In order to make a rough estimate of the implications of such a geometrical change, we for transverse diffusion consider the accessible diffusion-area per unit area in a lateral cross-section of SC. By using hexagonal prism Bricks and assuming its side to be $d$ gives the fraction $\sim 1.15\cdot s/d$ for small $s/d$. The same assumption gives $1.0\cdot s/d$ for the previously used cuboid Bricks. These similar values tell us that transverse diffusion in SC might be accurately modelled by cuboid Bricks. Lateral diffusion however is more complex to estimate since the previously one-dimensional offset ratio $\omega$ for a hexagonal prism system would become two-dimensional. As such the 1D model is inherently insufficient to capture the multidimensional conceptional characteristics of SC. To what degree this might affect the tortuosity is however unclear and needs to be investigated further. 

\section{Conclusions}
We have presented a compact (semi-)exact expression for the tortuosity in the Brick and Mortar model using impermeable Bricks. The derivation of this expression is based on the generalized Fick's law for steady-state mass diffusion. The presented method is advantageous to existing standard approaches due to its fewer approximations, generalizability, and its novel ability to capture anisotropic features of the Mortar which no other analytic model capture. The assumptions are extraneous for low/high off set ratios $\omega$ or small spacing between bricks, making it suitable for, among others, Stratum Corneum geometries. We furthermore introduce the concept of chemical conduction, after which we find Fick's, Ohm's and Fouriers law consistent in their constituent parts. Their common form simplifies conceptual understanding of diffusion as interchangeable parallels can be drawn between fields, and makes the main result of tortuosity in this work applicable for both electric, thermal, and mass diffusion.

\section*{Acknowledgements}
We want to acknowledge L'Or\'{e}al for funding; and H\r{a}kan Wennerstr\"{o}m for interesting discussions and useful comments. 



\bibliographystyle{elsarticle-num} 
\bibliography{manuscript}

\begin{thebibliography}{10}
\expandafter\ifx\csname url\endcsname\relax
  \def\url#1{\texttt{#1}}\fi
\expandafter\ifx\csname urlprefix\endcsname\relax\def\urlprefix{URL }\fi
\expandafter\ifx\csname href\endcsname\relax
  \def\href#1#2{#2} \def\path#1{#1}\fi

\bibitem{Onsager1932Irreversible}
L.~Onsager, R.~M. Fuoss, Irreversible processes in electrolytes. diffusion,
  conductance and viscous flow in arbitrary mixtures of strong electrolytes,
  The Journal of Physical Chemistry 36~(11) (1932) 2689--2778.
\newblock \href {http://arxiv.org/abs/https://doi.org/10.1021/j150341a001}
  {\path{arXiv:https://doi.org/10.1021/j150341a001}}, \href
  {http://dx.doi.org/10.1021/j150341a001} {\path{doi:10.1021/j150341a001}}.

\bibitem{dou2015transparent}
Y.~Dou, T.~Pan, S.~Xu, H.~Yan, J.~Han, M.~Wei, D.~G. Evans, X.~Duan,
  Transparent, ultrahigh-gas-barrier films with a brick--mortar--sand
  structure, Angewandte Chemie International Edition 54~(33) (2015) 9673--9678.

\bibitem{yao2010artificial}
H.-B. Yao, Z.-H. Tan, H.-Y. Fang, S.-H. Yu, Artificial nacre-like
  bionanocomposite films from the self-assembly of chitosan--montmorillonite
  hybrid building blocks, Angewandte Chemie International Edition 49~(52)
  (2010) 10127--10131.

\bibitem{Chai2014}
S.-H. Chai, P.~F. Fulvio, P.~C. Hillesheim, Z.-A. Qiao, S.~M. Mahurin, S.~Dai,
  \href{https://doi.org/10.1016/j.memsci.2014.05.044}{{\textquotedblleft}brick-and-mortar{\textquotedblright}
  synthesis of free-standing mesoporous carbon nanocomposite membranes as
  supports of room temperature ionic liquids for {CO}2-n2 separation}, Journal
  of Membrane Science 468 (2014) 73--80.
\newblock \href {http://dx.doi.org/10.1016/j.memsci.2014.05.044}
  {\path{doi:10.1016/j.memsci.2014.05.044}}.
\newline\urlprefix\url{https://doi.org/10.1016/j.memsci.2014.05.044}

\bibitem{Svagan2016solid}
A.~J. Svagan, J.-W. Benjamins, Z.~Al-Ansari, D.~B. Shalom, A.~Müllertz,
  L.~Wågberg, K.~Löbmann, Solid cellulose nanofiber based foams – towards
  facile design of sustained drug delivery systems, Journal of Controlled
  Release 244 (2016) 74 -- 82.
\newblock \href
  {http://dx.doi.org/https://doi.org/10.1016/j.jconrel.2016.11.009}
  {\path{doi:https://doi.org/10.1016/j.jconrel.2016.11.009}}.

\bibitem{Whitaker1986}
S.~Whitaker, \href{https://doi.org/10.1007/bf01036523}{Flow in porous media i:
  A theoretical derivation of darcy's law}, Transport in Porous Media 1~(1)
  (1986) 3--25.
\newblock \href {http://dx.doi.org/10.1007/bf01036523}
  {\path{doi:10.1007/bf01036523}}.
\newline\urlprefix\url{https://doi.org/10.1007/bf01036523}

\bibitem{OCHOATAPIA20071}
J.~A. Ochoa-Tapia, F.~J. Valdes-Parada, J.~Alvarez-Ramirez, A fractional-order
  darcy's law, Physica A: Statistical Mechanics and its Applications 374~(1)
  (2007) 1 -- 14.
\newblock \href {http://dx.doi.org/https://doi.org/10.1016/j.physa.2006.07.033}
  {\path{doi:https://doi.org/10.1016/j.physa.2006.07.033}}.

\bibitem{Heisig1996}
M.~Heisig, R.~Lieckfeldt, G.~Wittum, G.~Mazurkevich, G.~Lee,
  \href{https://doi.org/10.1023/a:1016048710880}{Non steady-state descriptions
  of drug permeation through stratum corneum. i. the biphasic brick-and-mortar
  model}, Pharmaceutical Research 13~(3) (1996) 421--426.
\newblock \href {http://dx.doi.org/10.1023/a:1016048710880}
  {\path{doi:10.1023/a:1016048710880}}.
\newline\urlprefix\url{https://doi.org/10.1023/a:1016048710880}

\bibitem{Hadgraft2011}
J.~Hadgraft, M.~E. Lane, \href{https://doi.org/10.1039/c0cp02943b}{Skin: the
  ultimate interface}, Physical Chemistry Chemical Physics 13~(12) (2011) 5215.
\newblock \href {http://dx.doi.org/10.1039/c0cp02943b}
  {\path{doi:10.1039/c0cp02943b}}.
\newline\urlprefix\url{https://doi.org/10.1039/c0cp02943b}

\bibitem{Walicka2018}
A.~Walicka, B.~Iwanowska-Chomiak,
  \href{https://doi.org/10.2478/ijame-2018-0055}{Drug diffusion transport
  through human skin}, International Journal of Applied Mechanics and
  Engineering 23~(4) (2018) 977--988.
\newblock \href {http://dx.doi.org/10.2478/ijame-2018-0055}
  {\path{doi:10.2478/ijame-2018-0055}}.
\newline\urlprefix\url{https://doi.org/10.2478/ijame-2018-0055}

\bibitem{bouwstra1991structural}
J.~A. Bouwstra, G.~S. Gooris, J.~A. van~der Spek, W.~Bras, Structural
  investigations of human stratum corneum by small-angle x-ray scattering,
  Journal of Investigative Dermatology 97~(6) (1991) 1005 -- 1012.
\newblock \href
  {http://dx.doi.org/https://doi.org/10.1111/1523-1747.ep12492217}
  {\path{doi:https://doi.org/10.1111/1523-1747.ep12492217}}.

\bibitem{cussler1988barrier}
E.~Cussler, S.~E. Hughes, W.~J. Ward~III, R.~Aris, Barrier membranes, Journal
  of membrane science 38~(2) (1988) 161--174.

\bibitem{johnson1997evaluation}
M.~E. Johnson, D.~Blankschtein, R.~Langer, Evaluation of solute permeation
  through the stratum corneum: lateral bilayer diffusion as the primary
  transport mechanism, Journal of pharmaceutical sciences 86~(10) (1997)
  1162--1172.

\bibitem{Griffiths1999Introduction}
D.~J. Griffiths, Introduction to electrodynamics., Prentice Hall, 1999.

\bibitem{Onsager1931Reciprocal}
L.~Onsager, Reciprocal relations in irreversible processes. i., Phys. Rev. 37
  (1931) 405--426.
\newblock \href {http://dx.doi.org/10.1103/PhysRev.37.405}
  {\path{doi:10.1103/PhysRev.37.405}}.

\bibitem{Onsager1953Fluctuations}
L.~Onsager, S.~Machlup, Fluctuations and irreversible processes, Phys. Rev. 91
  (1953) 1505--1512.
\newblock \href {http://dx.doi.org/10.1103/PhysRev.91.1505}
  {\path{doi:10.1103/PhysRev.91.1505}}.

\bibitem{crank1956mathematics}
J.~Crank, The mathematics of diffusion, Clarendon Press, Oxford, 1956.

\bibitem{Hearon1950}
J.~Z. Hearon, Some cellular diffusion problems based on onsager's
  generalization of fick's law, The bulletin of mathematical biophysics 12~(2)
  (1950) 135--159.
\newblock \href {http://dx.doi.org/10.1007/BF02478250}
  {\path{doi:10.1007/BF02478250}}.

\bibitem{evans1999colloidal}
D.~F. Evans, H.~Wennerstr\"{o}m, The colloidal domain : where physics,
  chemistry, biology, and technology meet., Advances in interfacial engineering
  series, Wiley-VCH, 1999.

\bibitem{Kocherginsky2016}
N.~Kocherginsky, M.~Gruebele,
  \href{https://doi.org/10.1073/pnas.1600866113}{Mechanical approach to
  chemical transport}, Proceedings of the National Academy of Sciences 113~(40)
  (2016) 11116--11121.
\newblock \href {http://dx.doi.org/10.1073/pnas.1600866113}
  {\path{doi:10.1073/pnas.1600866113}}.
\newline\urlprefix\url{https://doi.org/10.1073/pnas.1600866113}

\bibitem{kocherginsky2010mass}
N.~Kocherginsky, Mass transport and membrane separations: Universal description
  in terms of physicochemical potential and einstein's mobility, Chemical
  Engineering Science 65~(4) (2010) 1474--1489.

\bibitem{Wakeham1979}
W.~A. Wakeham, E.~A. Mason,
  \href{https://doi.org/10.1021/i160072a001}{Diffusion through multiperforate
  laminae}, Industrial {\&} Engineering Chemistry Fundamentals 18~(4) (1979)
  301--305.
\newblock \href {http://dx.doi.org/10.1021/i160072a001}
  {\path{doi:10.1021/i160072a001}}.
\newline\urlprefix\url{https://doi.org/10.1021/i160072a001}

\bibitem{Michaels1975Drug}
A.~S. Michaels, S.~K. Chandrasekaran, J.~E. Shaw, Drug permeation through human
  skin: Theory and invitro experimental measurement, AIChE Journal 21~(5)
  (1975) 985--996.
\newblock \href {http://dx.doi.org/10.1002/aic.690210522}
  {\path{doi:10.1002/aic.690210522}}.

\bibitem{Cussler1990}
E.~Cussler, \href{https://doi.org/10.1016/s0376-7388(00)85132-7}{Membranes
  containing selective flakes}, Journal of Membrane Science 52~(3) (1990)
  275--288.
\newblock \href {http://dx.doi.org/10.1016/s0376-7388(00)85132-7}
  {\path{doi:10.1016/s0376-7388(00)85132-7}}.
\newline\urlprefix\url{https://doi.org/10.1016/s0376-7388(00)85132-7}

\bibitem{LANGELIECKFELDT1992183}
R.~Lange-Lieckfeldt, G.~Lee, Use of a model lipid matrix to demonstrate the
  dependence of the stratum corneum's barrier properties on its internal
  geometry, Journal of Controlled Release 20~(3) (1992) 183 -- 194.
\newblock \href
  {http://dx.doi.org/https://doi.org/10.1016/0168-3659(92)90120-G}
  {\path{doi:https://doi.org/10.1016/0168-3659(92)90120-G}}.

\bibitem{KUSHNER20073236}
J.~Kushner, W.~Deen, D.~Blankschtein, R.~Langer, First‐principles,
  structure‐based transdermal transport model to evaluate lipid partition and
  diffusion coefficients of hydrophobic permeants solely from stratum corneum
  permeation experiments, Journal of Pharmaceutical Sciences 96~(12) (2007)
  3236 -- 3251.
\newblock \href {http://dx.doi.org/https://doi.org/10.1002/jps.20896}
  {\path{doi:https://doi.org/10.1002/jps.20896}}.

\bibitem{frasch2003steady}
H.~F. Frasch, A.~M. Barbero, Steady-state flux and lag time in the stratum
  corneum lipid pathway: Results from finite element models, Journal of
  pharmaceutical sciences 92~(11) (2003) 2196--2207.

\bibitem{edwards1994linear}
D.~A. Edwards, R.~Langer, A linear theory of transdermal transport phenomena,
  Journal of Pharmaceutical Sciences 83~(9) (1994) 1315--1334.
\newblock \href {http://dx.doi.org/10.1002/jps.2600830925}
  {\path{doi:10.1002/jps.2600830925}}.

\bibitem{yu2018geometrical}
F.~Yu, G.~B. Kasting, A geometrical model for diffusion of hydrophilic
  compounds in human stratum corneum, Mathematical Biosciences 300 (2018) 55 --
  63.
\newblock \href {http://dx.doi.org/https://doi.org/10.1016/j.mbs.2018.03.010}
  {\path{doi:https://doi.org/10.1016/j.mbs.2018.03.010}}.

\end{thebibliography}


\end{document}


\begin{frontmatter}



\title{Supplementary Information: \\Tortuosity in the Brick and Mortar model based on Chemical conduction}


\author{Bj\"orn Stenqvist}
\ead{bjorn.stenqvist@teokem.lu.se}
\author{Emma Sparr}
\address{Division of Physical Chemistry, Lund University, POB 124, SE-22100 Lund, Sweden}

\end{frontmatter}

Definitions and units of the used variables can, if not state here, be found in the main text of this work.

\section{Chemical conductivity}
The expressions for electric ($e$), thermal ($t$), and mass ($c$) transfer at steady-state has the same form,
\begin{subequations}\label{eq:laws}
\begin{align}
  \boldsymbol{j}_e &= -\sigma_e\boldsymbol{\nabla}\Phi,     && \text{Ohm's law} \\
  \boldsymbol{j}_t &= -\sigma_t\boldsymbol{\nabla} T,     && \text{Fourier's law} \\
  \boldsymbol{j}_m &= -\sigma_c\boldsymbol{\nabla}\mu, && \text{Generalized Fick's law}.\label{eq:laws_sub3}
\end{align}
\end{subequations}
Here $\boldsymbol{j}_e$ [C/m$^2$s] and $\boldsymbol{j}_t$ [J/m$^2$s] are the respective fluxes, $\Phi$ [V] is the electric potential, $T$ [K] the temperature, $\sigma_e$ [C/Vms] the electric conductivity, and $\sigma_t$ [J/Kms] the thermal conductivity. In the following we will show that by using the notion of chemical conductivity then Ohm's, Fourier's, and Generalized Fick's laws are consistent in terms of their interpretation and constituent parts. 

The mass flux can be expressed as\cite{kocherginsky2010mass}
\begin{equation}\label{eq:fick_first_exp}
\boldsymbol{j}_c = -U_mc\boldsymbol{\nabla} \mu.
\end{equation}
The mobility is defined as the steady-state velocity of a particle under the action of a unit force, and is thus inversely proportional to the viscosity in a fluid (based on Stokes equation) and to the friction coefficient\cite{kocherginsky2010mass}. After a comparison of Eq.~\ref{eq:laws_sub3} to this expression we find that
\begin{equation}\label{eq:masscond}
\sigma_c = U_mc.
\end{equation}
It has been recognised that for ideal systems $D_0c$ alone (or $\sim U_mc$ as we will later see) is sufficient to determine the effective self-diffusion coefficient\cite{jonsson1986self}. The derivation of this results is however straight-forward to generalize to non-ideal systems based on Eq.~\ref{eq:fick_first_exp}. By comparing Eq.~\ref{eq:fick_first_exp} to
\begin{equation}
\boldsymbol{j}_c = -D_0\boldsymbol{\nabla} c
\end{equation}
which serves as the definition of the diffusion coefficient\cite{crank1956mathematics,cussler1997diffusion}, we find that
\begin{equation}
D_0 = U_mc\frac{d\mu}{dc}
\end{equation}
which for ideal cases results in $D_0=U_mRT$, where $R$ [J/K mol] is the gas constant. The electric counterpart to Eq.~\ref{eq:masscond} is
\begin{equation}
\sigma_e = U_e\rho,
\end{equation}
where $U_e$ [Cs/kg] is the electric mobility and $\rho$ [C/m$^3$] the charge-density, and the thermal equivalent is
\begin{equation}
\sigma_t = U_t c_p
\end{equation}
where $U_t$ [Js/kg] is the thermal diffusivity (read $\sim$mobility) and $c_p$ [J/Km$^3$] the volumetric heat capacity. We now acknowledge the similarities between the laws in Eq.~\ref{eq:laws}, both in form and coupling to other entities. Thus by using the definition of the chemical conductivity (for simplicity in one dimension)
\begin{equation}
\sigma_c \equiv -\frac{j_m}{\nabla\mu},
\end{equation}
we have consistency between Ohm's, Fourier's, and the generalized Fick's law. The definition, or rather its macroscopic equivalent
\begin{equation}
\sigma_c = -j_m\frac{l}{\Delta\mu},
\end{equation}
where $l$ is the length over which the difference $\Delta\mu$ is measured, makes it possible to obtain a value of the chemical conductivity from experiments.

Finally we note that since the generalized Fick's law in Eq.~\ref{eq:laws} has the same form as the ideal Fick's law, we can trivially reuse past solutions to the ideal Fick's law simply by replacing the diffusion coefficient $D_0$ by the chemical conductivity $\sigma_c$, and the concentration $c$ by the chemical potential $\mu$. Also, in Sec.~\ref{sec:alternative} we derive an alternative form of the generalized Fick's law based on the ($\sim$) activity coefficient which might be more appropriate for some problems.

\section{Alternative formalism}\label{sec:alternative}
The chemical conductivity is a product of the mobility and the concentration, and thus these parameters needs to be known (at least their product) in advance if using the procedure in the main text. For scenarios when this product is not known, we here present an alternative formalism which instead relies on knowing the mobility $U_m$ and the activity coefficient $\gamma$ of the system. As in the main text we start by stating a generalized form of Fick's first law\cite{kocherginsky2010mass} which can be written as
\begin{equation}\label{eq:fick_first_exp_SI}
\boldsymbol{j}_c = -U_mc\boldsymbol{\nabla} \mu.
\end{equation}
By using the definitions\cite{mills1993quantities} of the activity $a$ [unitless] and the activity coefficient $\gamma$ [unitless]
\begin{equation}\label{eq:activity_def_SI}
\exp\left(\frac{\mu-\mu^{\circ}}{RT} \right) \equiv a \equiv \gamma\frac{c}{c^{\circ}},
\end{equation}
where $\mu^{\circ}$ [J/mol] is the standard chemical potential, and $c^{\circ}$ [kg/m$^3$] the standard concentration, we can rearrange the generalized form of Fick's first law as
\begin{equation}\label{eq:fick_gen_SI}
\boldsymbol{j}_c = -\frac{D_0}{\gamma}\boldsymbol{\nabla}\left(\gamma c\right)
\end{equation}
where we have defined $D$ as
\begin{equation}\label{eq:diff_def_SI}
D \equiv \frac{D_0}{\gamma} = \frac{U_mRT}{\gamma}
\end{equation}
and assumed constant $\exp\left(\mu^{\circ}/RT\right)/c^{\circ}$. We note that $D$ has the same unit as the diffusion coefficient and that in the ideal limit, i.e. $\gamma\to 1$, we have $D\to D_0$. Therefore we denote Eq.~\ref{eq:diff_def_SI} as the definition of the non-ideal diffusion coefficient. 

Now we follow the procedure in the main text for retrieving the tortuosity, yet here we base the derivation on Eq.~\ref{eq:fick_gen_SI}. For simplicity we in the following use the notation $\Tilde{a}=\gamma c$ since the product is closely related to the activity $a$. The solution to the 1D steady-state problem is then
\begin{equation}\label{eq:fick_sol_SI}
\Tilde{a}(z) = \Tilde{a}(z_0) + \lambda(z)\Delta\Tilde{a},
\end{equation}
where $\Delta\Tilde{a}=\Tilde{a}(z_f) - \Tilde{a}(z_0)$ and
\begin{equation}
\lambda(z) = \frac{\int_{z_0}^zD^{-1}(z^{\prime})dz^{\prime}}{\int_{z_0}^{z_f}D^{-1}(z^{\prime})dz^{\prime}},
\end{equation}
which gives the flux as
\begin{equation}\label{eq:gen_flux_SI}
j_m = -\frac{\Delta\Tilde{a}}{\int_{z_0}^{z_f}D^{-1}(z)dz}.
\end{equation}
By using the definition
\begin{equation}\label{eq:resistance_SI}
R = \frac{\Delta\Tilde{a}}{\dot{m}}
\end{equation}
for the resistance instead of Eq.~14 (main text), the final expression for the tortuosity then becomes
\begin{equation}\label{eq:tau_final_SI}
\tau_{\mu} = 1+\frac{d}{s} + \frac{L^2\omega}{g\left(g+t\frac{N}{N-1}\right)(1+\omega)^2}\frac{D_{\perp}}{D_{\parallel}}.
\end{equation}
The difference between Eq.~\ref{eq:tau_final_SI} and Eq.~23 (main text), is only in the last term which scales with $D_{\perp}/D_{\parallel}$ compared to $\sigma_{\perp}/\sigma_{\parallel}$, and the corresponding assumptions are on knowing these entities. An important observation is however also the definition of resistance which now is based on Eq.~\ref{eq:resistance_SI}.

\section{General expression for the resistance}
In the main work we for simplicity used a model which assumes constant lateral chemical conduction and constant transverse chemical conduction. Here we present a general procedure which makes no such assumptions, and therefore is applicable for non-constant conductivity $\sigma_c(x,z)$. 

The resistance over a line-segment $\mathcal{C}$ is
\begin{equation}
R = \frac{1}{A}\int_{\mathcal{C}}\sigma_c^{-1}(\boldsymbol{r})ds
\end{equation}
which gives the effective resistance over parallel such curves as
\begin{equation}
\frac{1}{R^*} = \sum_i\frac{1}{R_i} = \sum_i\frac{A_i}{\int_{\mathcal{C}_i}\sigma_c^{-1}(\boldsymbol{r})ds}.
\end{equation}
The above expression can then be simplified to
\begin{equation}
R^* = \frac{\prod\limits_{k}\int{\mathcal{C}_k}\sigma_c^{-1}(\boldsymbol{r})ds}{\sum\limits_iA_i\prod\limits_{k\ne i}\int_{\mathcal{C}_k}\sigma_c^{-1}(\boldsymbol{r})ds}.
\end{equation}
For a Brick and Mortar model we thus have
\begin{equation}
R_N = \overbrace{\frac{\int_0^h\sigma_c^{-1}(x(z),z)dz}{A_t}}^{\text{$R_t$ -  Transverse resistance}} + \overbrace{\sum_{n=1}^{N-1}\frac{\prod\limits_{k\in\{\leftarrow,\rightarrow\}}\int_{x_n}^{x_n+l_k}\sigma_c^{-1}(x,z_n)dx}{A_l\sum\limits_{i\in\{\leftarrow,\rightarrow\}}\prod\limits_{k\ne i}\int_{x_n}^{x_n+l_k}\sigma_c^{-1}(x,z_n)dx}}^{\text{$R_l$ - Lateral resistance}}
\end{equation}
where $x_n$ is the first reached crossing-point in layer $n$ for a diffusing particle,
\begin{equation}
z_n = \left(t+g\right)\left(n-1\right) + \left(t+\frac{g}{2}\right),
\end{equation}
and
\begin{equation}
x(z) = \begin{cases}
0 &\text{$z\in$ Odd numbered layer}\\
l_{\rightarrow} (\text{or equivalently } -l_{\leftarrow}) &\text{$z\in$ Even numbered layer}.
\end{cases}
\end{equation}

\section{The assumptions of the procedure}
\begin{figure}[!ht]
  \caption{An enhanced section of Fig.~1a in the main text, where we illustrate the approximated diffusion-paths by the dotted line and the colored segments. }
  \centering
    \includegraphics[width=0.8\columnwidth]{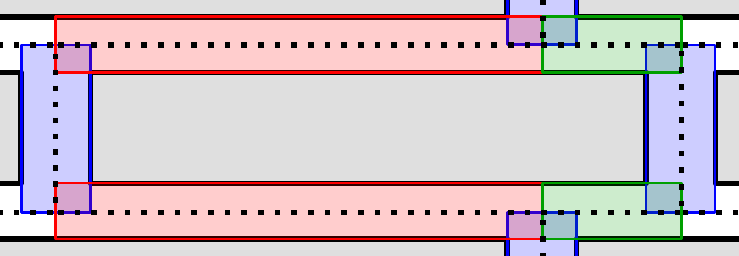}
    \label{fig:assumptions}
\end{figure}

The approximations in this work all stems from the fact that we assume a 1D solution to be accurate. Within this approximation lies two important features to consider. Firstly, the approximated diffusion volume will effectively \emph{double-count} some volumes, see the overlap between colored segments in Fig.~\ref{fig:assumptions}, and it will thus differ from the actual diffusion volume. Secondly, the properties of the diffusing particles is assumed to be independent of any coordinate perpendicular to the straight diffusion-path, see the dotted lines in Fig.~\ref{fig:assumptions}. Thus, for example if in a red segment in Fig.~\ref{fig:assumptions}, a diffusing particle will not be affected by a transverse perturbation, that is a \emph{uniform} transverse distribution.

The first feature of double-counting is nullified by using either $g\to 0$, $s\to 0$ or $N=0$. Then the approximated diffusion volume equals the actual diffusion volume. Even if not zero, just by decreasing $g$ and $s$ one will decrease the over-counted volume. Moreover, for $\omega=\infty$ the effect will not occur either since in steady-state for this case there is no lateral diffusion. The second feature about uniformity will also be nullified by using either $g\to 0$, $s\to 0$, $N=0$, or $\omega=\infty$. Here however the impact for other values of $g$, $s$, $N$, and $\omega$ is harder to evaluate and for many scenarios, intermediate values might still be fairly approximated by a uniform perpendicular (to the diffusion-path) distribution of properties.

\section{Linear chemical conductivity}
Now we assume that $\sigma_c$ is a linear function of $z$ as
\begin{equation}
\frac{\sigma_c(z)}{\sigma_0} = \frac{2a}{h}z + (1-a).
\end{equation}
This function thus gives a ``mean'' value of $\sigma_0$ over the interval $z\in[0,h]$, with end-points values of $(1\pm a)\sigma_0$. The transverse resistance then becomes
\begin{equation}
R_t = \frac{h}{A_t\sigma_0}\frac{1}{2a}\ln\left(\frac{1+a}{1-a}\right)
\end{equation}
and the lateral resistance
\begin{equation}
R_l = \frac{l_{\leftarrow}l_{\rightarrow}h}{A_lP\sigma_0}\sum_{n=1}^{N-1}\frac{1}{C_1n+C_2}
\end{equation}
where
\begin{equation}
C_1 = 2a(t+g)
\end{equation}
and
\begin{equation}
C_2 = h(1-a)-ag.
\end{equation}
For $N=0$ and constant $h$ and $L$ we have
\begin{equation}
R_0 = \frac{h}{L_DL\sigma_0}\frac{1}{2a}\ln\left(\frac{1+a}{1-a}\right)
\end{equation}
which finally gives, assuming $A_l=L_Dg$ and $A_t=L_Ds$ like before,
\begin{equation}\label{eq:tau_a_SI}
\tau_{\mu} = 1+\frac{d}{s} + \frac{L^2\omega}{g(g+t\frac{N}{N-1})(1+\omega)^2}\left[\frac{1}{(N-1)}\sum_{n=1}^{N-1}\frac{h}{C_1n+C_2}\right]\frac{2a}{\ln\left(\frac{1+a}{1-a}\right)}.
\end{equation}
In the limit $a\to 0$ this expression collapses to Eq.~22 in the main text (using $\sigma_{\perp}=\sigma_{\parallel}$).

\section{Further comparisons between $\tau_{\mu}$ and $\tau_c$}
\begin{figure}[!ht]
    \centering
    \begin{subfigure}[b]{0.5\columnwidth}
        \includegraphics[width=\textwidth]{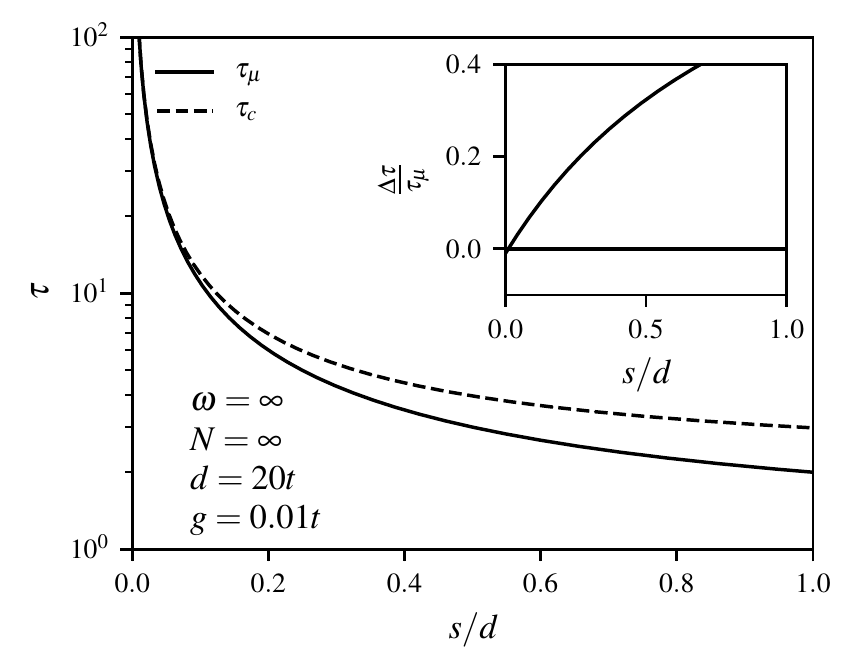}
        \caption{}
        \label{fig:tau_compare_s_SI}
    \end{subfigure}
    ~ 
    \begin{subfigure}[b]{0.5\columnwidth}
        \includegraphics[width=\textwidth]{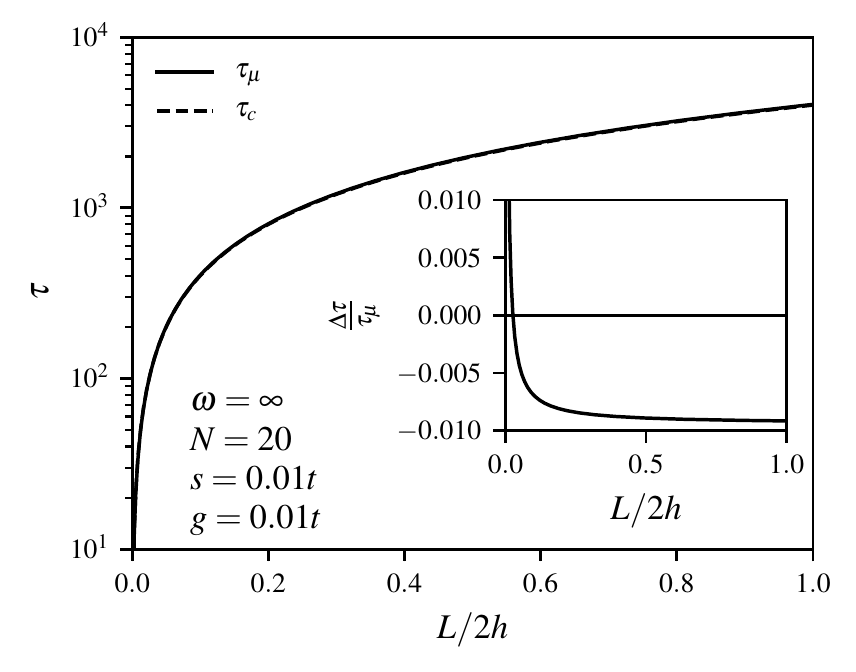}
        \caption{}
        \label{fig:tau_compare_d_SI}
    \end{subfigure}
    ~ 
    \begin{subfigure}[b]{0.5\columnwidth}
        \includegraphics[width=\textwidth]{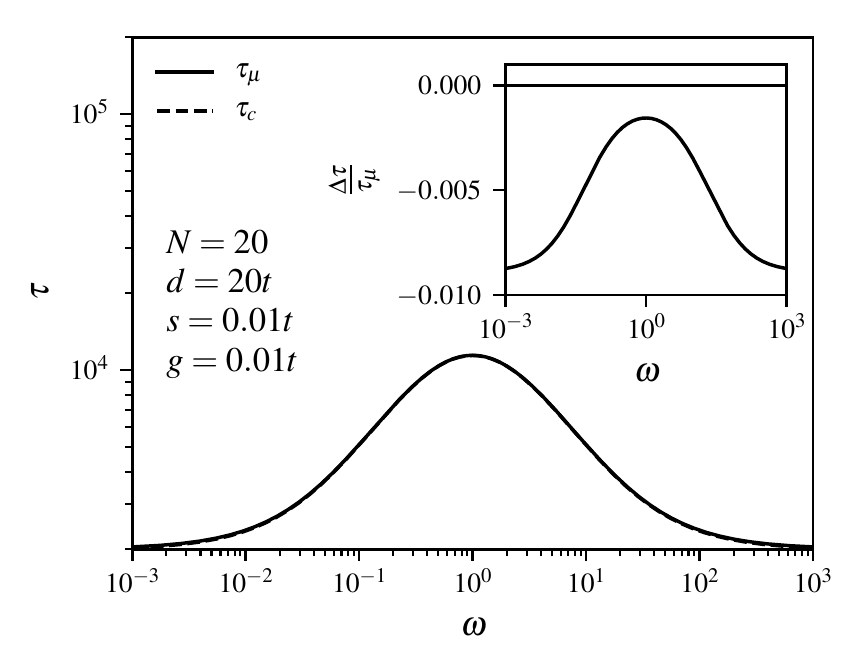}
        \caption{}
        \label{fig:tau_compare_o_SI}
    \end{subfigure}
    \caption{Tortuosity as a functions of (a) $s/d$ where $\tau_c$ is valid for $s/d\ll 1$ and $\tau_{\mu}$ for small $s$, (b) $L/2h$ where $\tau_c$ is valid for $L/2h\ll 1$, and (c) $\omega$ where $\tau_{\mu}$ is exact for $\omega=\infty$ (or equivalently $\omega=0$). The insets show the difference between the two approaches $\Delta\tau=\tau_c-\tau_{\mu}$ divided by $\tau_{\mu}$. Note in (b) the lower limit of $L/2h=438^{-1}$ for $\tau_{\mu}$ at which $d=0$ and we thus have a homogeneous membrane -- that is unit tortuosity.}\label{fig:tau_compare_SI}
\end{figure}

In Fig.~\ref{fig:tau_compare_SI} we present the same content as in Fig.~2, except that $g$ and/or $s$ has been scaled by a tenth. All comparisons in this section will be to the results presented in Fig.~2 in the main work.

In Fig.~\ref{fig:tau_compare_s_SI} we effectively see a shift of the error such that for small $s/d$ the expressions for the tortuosities are (roughly) equal, and in the limit $g\to 0$ and $s\to 0$ (for the presented scenario) they are indeed equal. For large $s/d$ the error does however increase and for $s/d\to\infty$ (not shown) $\Delta\tau/\tau_{\mu}=1/(1+g/t) \approx 1$. Interestingly we see that for small $g/t$, i.e. comparably small vertical spacing between Bricks to the thickness of the Bricks, the difference is $\sim 100\%$ whereas for large $g/t$ the difference is $\sim 0\%$.

Fig.~\ref{fig:tau_compare_d_SI} shows the same infinite difference as $L/2h\to 0$, yet for larger values it has basically been scaled by a tenth, which is what both $g$ and $s$ has been scaled by compared to Fig.~2. However, in this limit there is no $s$-dependence, but we rather have
\begin{equation}
\lim_{\frac{P}{2h}\to \infty}\left(\frac{\Delta\tau}{\tau_{\mu}}\right) =-\frac{1}{\left(1+\frac{t}{g}\frac{N}{N-1}\right)}
\end{equation}
where we note that for reasonable large $N$, the difference arise from the $t/g$-ratio. Thus, for $g\ll t$ (in this scenario) we have $\tau_{\mu}\approx\tau_c$ whereas for other values there is a significant difference.

Again we, in Fig.~\ref{fig:tau_compare_o_SI}, basically see a reduction of the difference by a tenth. Like before, but now independent of any parameter-choice like $d$ or $N$, we in the limit of small $s$ get
\begin{equation}
\lim_{s\to 0}\left(\frac{\Delta\tau}{\tau_{\mu}}\right)  = -\frac{1}{\left(1+\frac{t}{g}\frac{N}{N-1}\right)}.
\end{equation}
Thus again we acknowledge the importance of the $t/g$-ratio when it comes to the difference between $\tau_{\mu}$ and $\tau_c$.

\section{Vertical Bricks}
\begin{figure}[!ht]
  \caption{Tortuosity for vertically stretched Bricks as a functions of $s/d$ where $\tau_c$ is valid for $s/d\ll 1$ and $\tau_{\mu}$ for small $s$. The insets show the difference between the two approaches $\Delta\tau=\tau_c-\tau_{\mu}$ divided by $\tau_{\mu}$.}
  \centering
    \includegraphics[width=0.8\columnwidth]{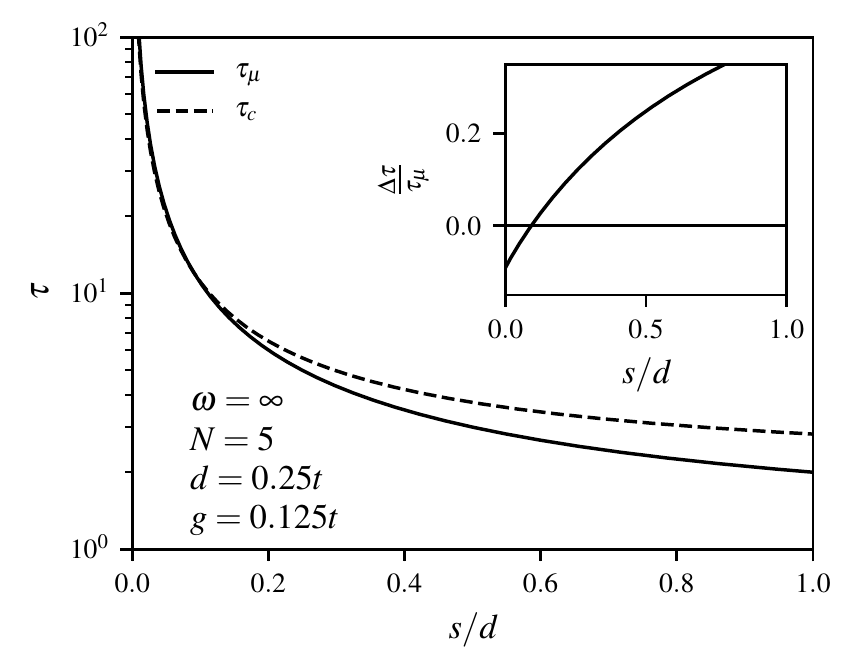}
    \label{fig:verbricks}
\end{figure}

In Fig.~\ref{fig:verbricks} we present the case when the Bricks are vertically stretched, as compared to the so far compared horizontally alignment. Compared to the corresponding case in the main text, Fig.~2a, the tortuosity is remarkably similar. Note however that the used values of $N$, $d$, and $g$ are slightly different.

\bibliographystyle{elsarticle-num} 
\bibliography{manuscript}
